\documentclass[twocolumn,showpacs,preprintnumbers,amsmath,amssymb]{revtex4}

\usepackage{graphicx}
\usepackage{dcolumn}
\usepackage{bm}
\usepackage{amsmath,dsfont}
\usepackage{amssymb,amsthm,amscd, amsbsy, array}
\input amssym.def
\input amssym.tex

\newcommand{\half}{{\scriptstyle{\frac{1}{2}}}}
\def\2{{\half}}

\newcommand{\vP}{{\bm P}}

\newcommand{\vp}{{\bm p}}
\def\vE{{\bm E}}

\def\br{{\bm{r}}}
\def\beq{\begin{equation}}
\def\eeq{\end{equation}}
\def\beqa{\begin{eqnarray}}
\def\eeqa{\end{eqnarray}}
\def\nn{\nonumber}
\def\barray{\left(\begin{array}}
\def\earray{\end{array}\right)}
\def\barraynb{\begin{array}}
\def\earraynb{\end{array}}

\def\smallover#1/#2{\hbox{$\textstyle\frac{#1}{#2}$}} %

\def\vx{{\bm{x}}}

\def\vX{{\bm{X}}}

\newcommand{\vK}{{\bm K}}

\def\half{\frac{1}{2}}
\def\ben{\begin{equation}}
\def\een{\end{equation}}
\def\bea{\begin{eqnarray}}
\def\eea{\end{eqnarray}} 

\def \nn{\nonumber}

\begin{document}

\preprint{arXiv: 1202.5081v2}

\title{Exotic  Hill problem~: Hall motions
and symmetries} 

\author{P. M. Zhang$^{1}$\footnote{email:zhpm@impcas.ac.cn},
P. A. Horvathy$^{1,2}$\footnote{email:horvathy@lmpt.univ-tours.fr}}

\affiliation{$^1$Institute of Modern Physics (CAS), Chinese Academy of Sciences
\\
Lanzhou, China \\
$^2$Laboratoire de Math\'ematiques et de Physique
Th\'eorique,
Universit\'e de Tours.
Tours, France}
 
\date{\today}

\begin{abstract}
Our previous study of a system of bodies assumed to move along almost circular orbits around a central mass, approximately described by Hill's equations, is
extended to ``exotic'' [alias non-commutative] particles. 
For a certain critical value of the angular velocity, the only allowed motions follow the Hall law. Translations and generalized   boosts span two independent Heisenberg algebras with different central parameters. In the critical case, the symmetry 
reduces to a single Heisenberg algebra.
\end{abstract}

\pacs{
11.30.-j, 	
02.40.Yy, 	
02.20.Sv, 	
96.12.De, 	
\\[8pt]
%
%
}

\maketitle


\section{Introduction}

The Hill problem arises as an approximation for nearly circular trajectories to 
Newton's gravitational equations written in rotating coordinates for  bodies moving around a central mass.
The original example is provided by
the Moon-Earth-Sun system \cite{Hill}. Later,
Hill's equations have also been applied  
to stellar dynamics \cite{Bok,Heggie}, with a ``star cluster'' replacing Moon and Earth, and the ``Galactic Center'' playing the role of the Sun. 

For the center-of-mass $\br=\sum_am_a\br_a/m$ where  $m=\sum_am_a,$
the inter-particle interactions cancel, leaving us,
in the planar case, with the  equations \cite{ZGH}, 
\beq
\begin{array}{llll}
\ddot{x}&-\,2\omega\dot{y}-3\omega^2x&=&0,
\\[4pt]
\ddot{y}&+\,2\omega\dot{x}&=&0,
\end{array}
\label{HillSymmeqs} \,
\eeq
where $\omega$ is the angular velocity $\omega^2=GM/R^3$ for
a circular Keplerian trajectory with radius  $R$. Note here the ``only-$x$''
anisotropic oscillator term, which is the remnant  
of the centrifugal and Newtonian forces under linear approximation.

The general solution of eqn. (\ref{HillSymmeqs}) is a combination of simply ``Hall'' motions of a guiding center, combined with elliptic motions around the guiding center \cite{ZGH}.

The decomposability into center-of-mass and relative coordinates  relies on having an Abelian symmetry made
of (generalized) translations and boosts,
characteristic for Galilean symmetry \cite{SSD,GalKohn}. 
These symmetries are also given by the solutions of eqn.  (\ref{HillSymmeqs}); 
the conserved quantities associated with them span two copies of Heisenberg algebras with  central parameters 
$\mp2/m\omega$ \cite{ZGH}.

The velocity-dependent terms in eqn. (\ref{HillSymmeqs}) 
come from the Coriolis force, and are analogous to a uniform magnetic field. Whence the similarity to the Landau problem and to the Hall effect.

On the other hand, some insight  to the
Hall effect could be gained by assuming that the charged particles admit an
additional ``exotic'' structure,
 which makes the coordinates $x$
and $y$ non-commuting  \cite{DHexo}.

In this Note we generalize Hill's equations
to exotic particles, providing us with a combination 
of the Hill results in \cite{ZGH} with
the Hall-type behavior found before in the non-commutative
Landau problem \cite{NCLandau}.

Our main result is that, 
for the critical value  $\omega=\omega_c$ of the rotation i.e.
of the radius, 
the only allowed motions are those which satisfy the Hall law.

 Our basic tool is the generalization to the exotic Hill case of the chiral decomposition of Ref.
\cite{AGKP,ZH-chiral}.

\section{Exotic Hill solutions}

The planar Hill equations (\ref{HillSymmeqs})
derive from the  
 symplectic form and Hamiltonian \cite{ZGH}
\beq\begin{array}{lll}
\Omega_0&=&dp^x\wedge dx+dp^y\wedge dy
+2m\omega\, dx\wedge dy,
\\[6pt]
H&=&\displaystyle\frac{\vp^2}{2m}+\displaystyle\half kx^2,
\quad
k=-3m\omega^2\,.
\end{array}
\label{0sympHam}
\eeq
For an exotic particle, the symplectic form is
modified as
\begin{equation}
\Omega=\Omega_0+\frac{\theta}{2}\varepsilon^{ij}dp^i\wedge dp^j,
\label{exosymp}
\end{equation}
while the Hamiltonian is unchanged \cite{DHexo}. The 
constant $\theta$ is the \emph{non-commutative parameter},
as justified by the associated Poisson brackets 
\begin{equation}
\left\{ x^i,x^j\right\}=\frac{\theta}{\Delta}\varepsilon
^{ij},\;\;\left\{ x^i,p^j\right\}=\frac{\delta^{ij}}{\Delta}%
,\;\;\;\left\{p^i,p^j\right\}=\frac{2m\omega}{\Delta}
\varepsilon^{ij},\quad
\label{exoPB}
\end{equation}
where 
\beq
\Delta=
1-2m\omega\theta
\label{Delta}
\eeq
is the [square-root of the] determinant of the ``exotic''
symplectic form (\ref{exosymp}).

Hamilton's equations could be worked out and
the solution  found.
We find it, however, more convenient to
deduce them in a smarter way, presented in the next Section.

\section{Chiral decomposition}

Following \cite{AGKP,ZH-chiral,ZGH}, we introduce  chiral coordinates $x^i=X_{+}^i+X_{-}^i$,
\beq
p^1=\alpha _{+}X_{+}^2+\alpha _{-}X_{-}^2,
\quad
p^2=-\beta
_{+}X_{+}^1-\beta_{-}X_{-}^1,
\eeq
where the coefficients
$\alpha_\pm$ and $\beta_\pm$ are  determined
from the requirement that both the symplectic 
form and the Hamiltonian should split.
Then the calculation analogous to the one in \cite{ZGH}
yields 
\beq\barraynb{lll}
\Omega&=&
-\displaystyle
\frac{\Delta}{\Gamma}\,
\frac{m\omega}{2}\,dX_{+}^1\wedge dX_{+}^2
+
\displaystyle\frac{m\omega}{2}
dX_{-}^1\wedge dX_{-}^2,
\\[16pt]
H&=&\displaystyle\frac{m\omega^2}{2}\left(
X_{+}^1X_{+}^1
+\frac{1}{4\Gamma^2}X_{+}^2X_{+}^2
\right)
-
\displaystyle\frac{3m\omega^2}{8}
X_{-}^1X_{-}^1,\quad
\earraynb
\label{decomposedHill}
\eeq
where
\beq
\Gamma=
1-\smallover{3}/{2}\theta m\omega.
\eeq 

For $\theta=0$ we have $\Delta=\Gamma=1$, and the commutative Hill case \cite{ZGH} is recovered.

For $k=0$ the oscillator term is switched off, and
the system 
reduces to the purely magnetic non-commutative Landau problem 
discussed in \cite{AGKP,ZH-chiral}.

\section{Motions}

\begin{figure} 
\begin{center}
\includegraphics[scale=.35]{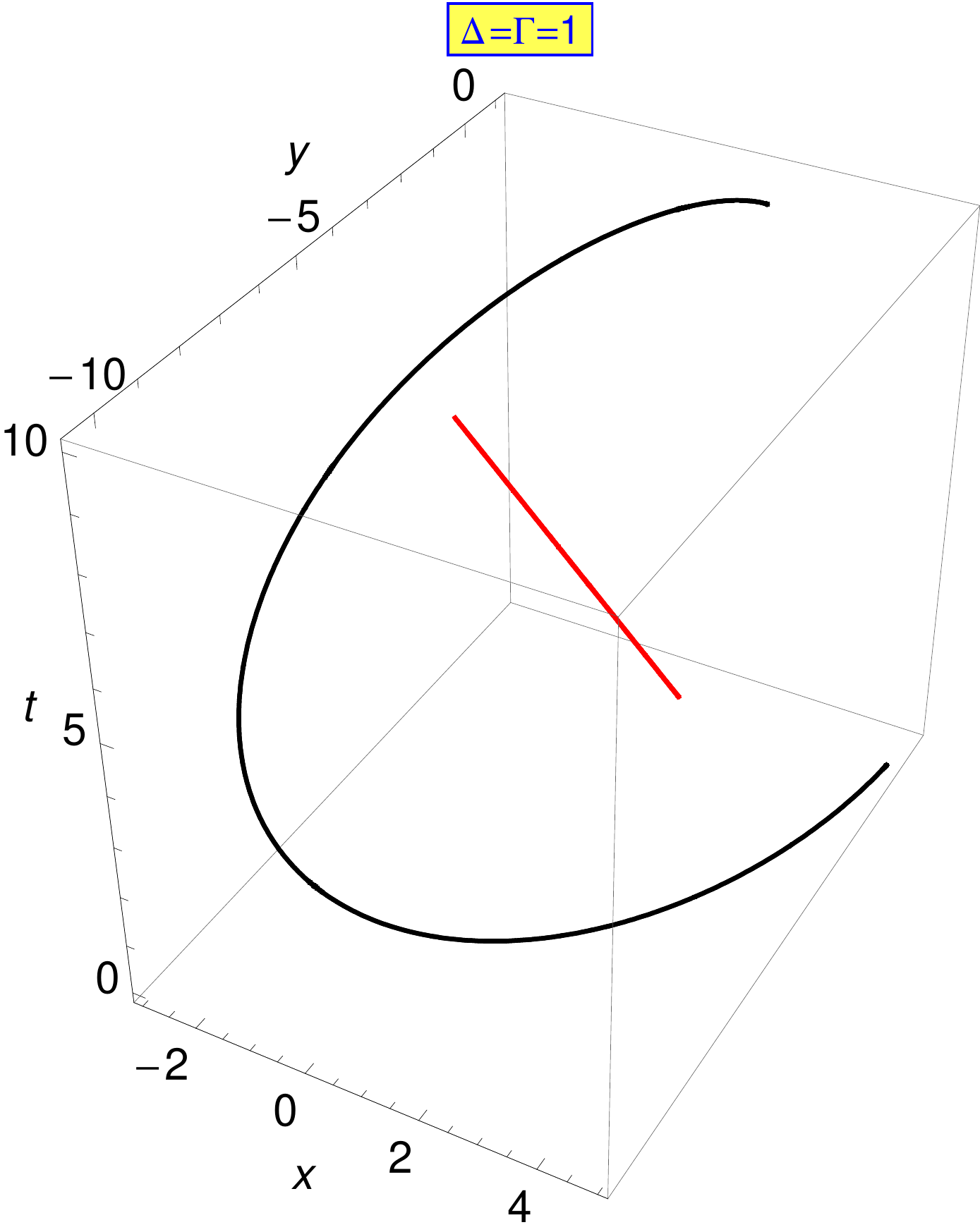}\\[10pt]
\includegraphics[scale=.35]{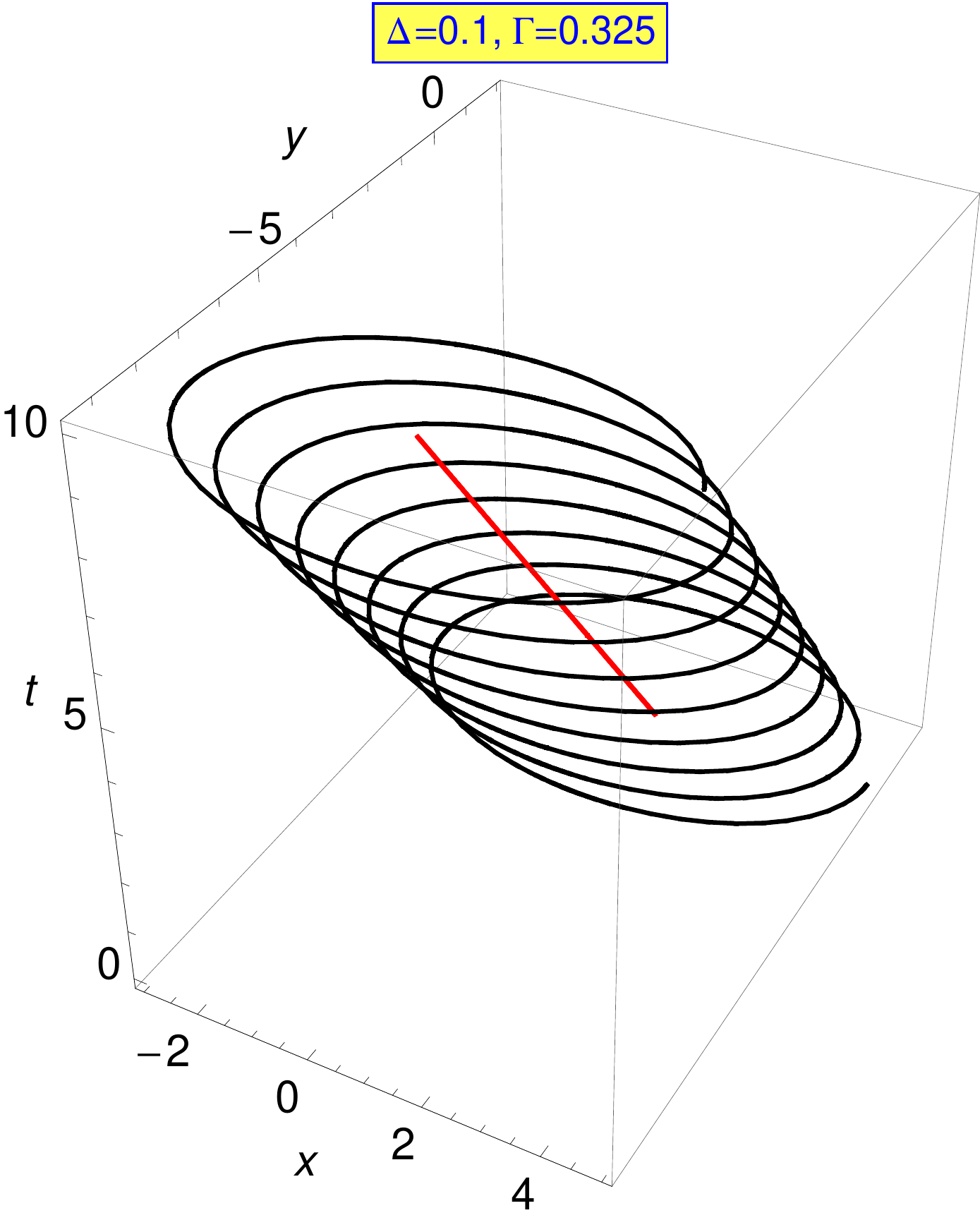}\\[10pt]
\includegraphics[scale=.35]{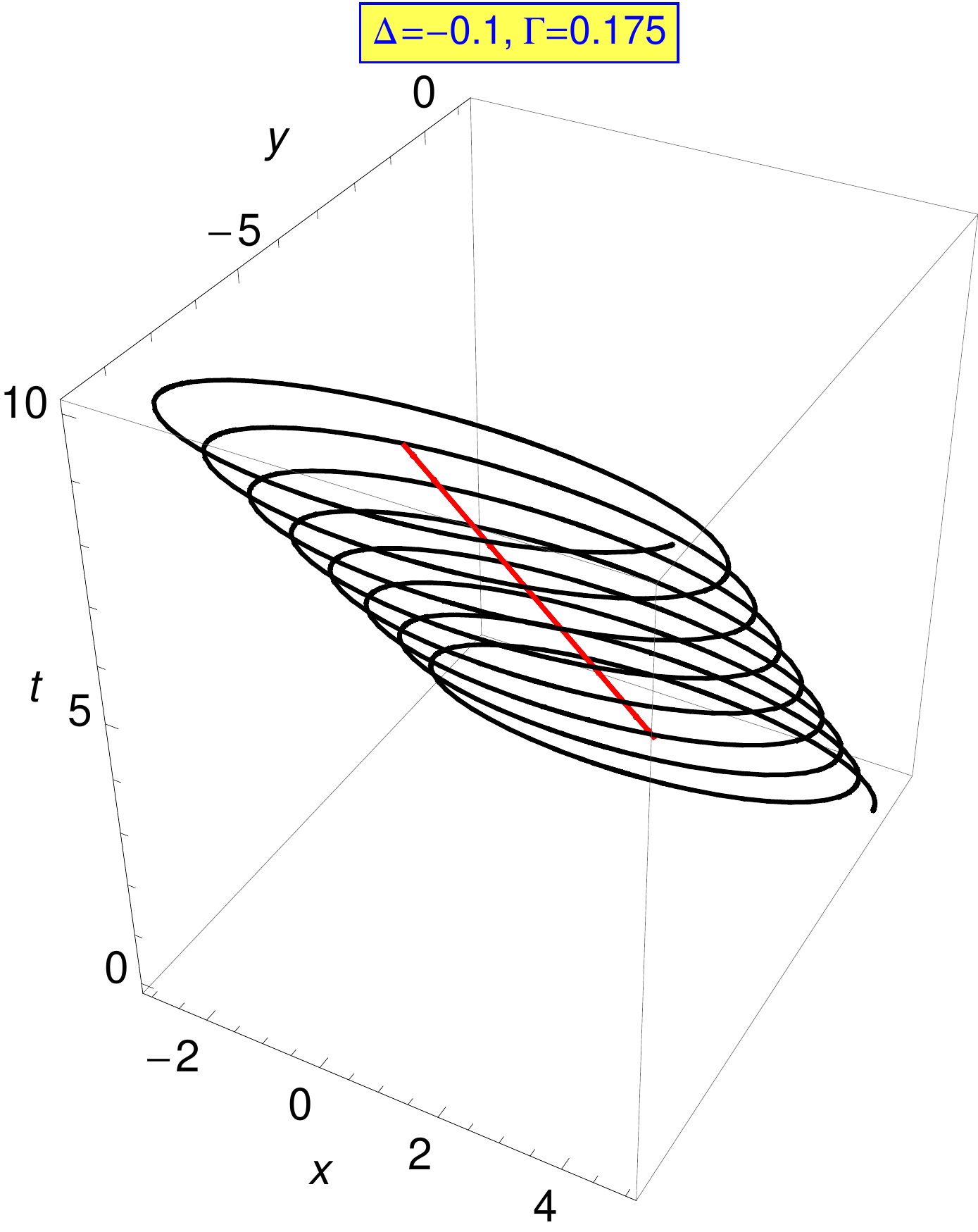}
\vspace{-7mm}
\end{center}
\caption{\it Trajectories for the exotic Hill equations unfolded into time (the vertical coordinate).
When the critical value $\Delta=0$ is approached, 
``non-Hall'' trajectories speed up,
while the guiding center (in red) moves following the Hall law. The latter are the only allowed motions when $\Delta=0$.
After crossing the critical value, the sense of rotation
is reversed. }
\label{k-figs}
\end{figure}

The decomposition
(\ref{decomposedHill}) implies the Poisson brackets
\begin{equation}
\left\{X_{+}^1,X_{+}^2\right\}=\displaystyle\frac{\Gamma}{\Delta}\,
\frac{2}{m\omega},
\qquad
\left\{X_{-}^1,X_{-}^2\right\}=-\displaystyle\frac{2}{m\omega}
\end{equation}
completed with $\{X_{+}^i,X_{-}^j\}=0$,
providing us with separated  equations of motion,
\beq\barraynb{lllllll}
\dot{X}_{-}^1&=&0\,,
&\null\qquad
&\dot{X}_{-}^2&=&-\displaystyle\frac{3}{2}\omega X_{-}^1\,,
\\[12pt]
\Gamma\Delta\,\dot{X}_{+}^1&=&\displaystyle\frac{\omega}{2}
X_{+}^2\,,
&\null\qquad
&\Delta\,\dot{X}_{+}^2&=&-2\Gamma
\omega\, X_{+}^1\,.
\earraynb
\label{X+-eq}
\eeq
Off the critical case, $\Delta\neq0$, the solution
is therefore
\beqa
X_{-}^1(t)&=&x_0,\;\;\;\;X_{-}^2(t)=-\frac{3}{2}\omega x_0t+y_0,
\label{X-Hall}
\\[10pt]
X_{+}^1(t)&=&\frac{A}{\omega ^{*}}\sin\omega^{*}t-\frac{B}{\omega^{*}}\cos\omega^{*}t,
\quad
\omega^*=\frac{\omega}{\Delta},
\\[6pt]
X_{+}^2(t)&=&\Gamma\left(2\frac{A}{\omega^{*}}\cos\omega
^{*}t+2\frac{B}{\omega^{*}}\sin\omega^{*}t\right),
\eeqa
$\vX_-(t)$ performs, hence, for all  $\Delta\neq0$ and $\Gamma$, a
simple uniform translational motion,
and can be identified with the \emph{guiding center coordinate}.
 $\vX_+(t)$, which moves along a flattened ellipses, describes instead motions about the guiding center.
 Note that  the $\vX_+$ dynamics depends on the non-commutative parameter $\theta$ through $\omega^*$,
 while the guiding center dynamics is  $\theta$-independent.
In terms of the original coordinates, the motion is
\beq\barraynb{lll}
x(t)&=&\displaystyle\frac{A}{\omega^*}\sin\omega^*t
-\displaystyle\frac{B}{\omega^*}\cos\omega^*t
+x_0,
\\[16pt]
y(t)&=&\Gamma\left(2\displaystyle\frac{A}{\omega^{*}}\cos\omega
^{*}t+2\displaystyle\frac{B}{\omega^{*}}\sin\omega^{*}t\right)
-\displaystyle\frac{3}{2}\omega\,x_0t+y_0\,.\quad
\earraynb
\label{gensol}
\eeq
see Fig. \ref{k-figs}.
%

\begin{figure} 
\begin{center}
\includegraphics[scale=.33]{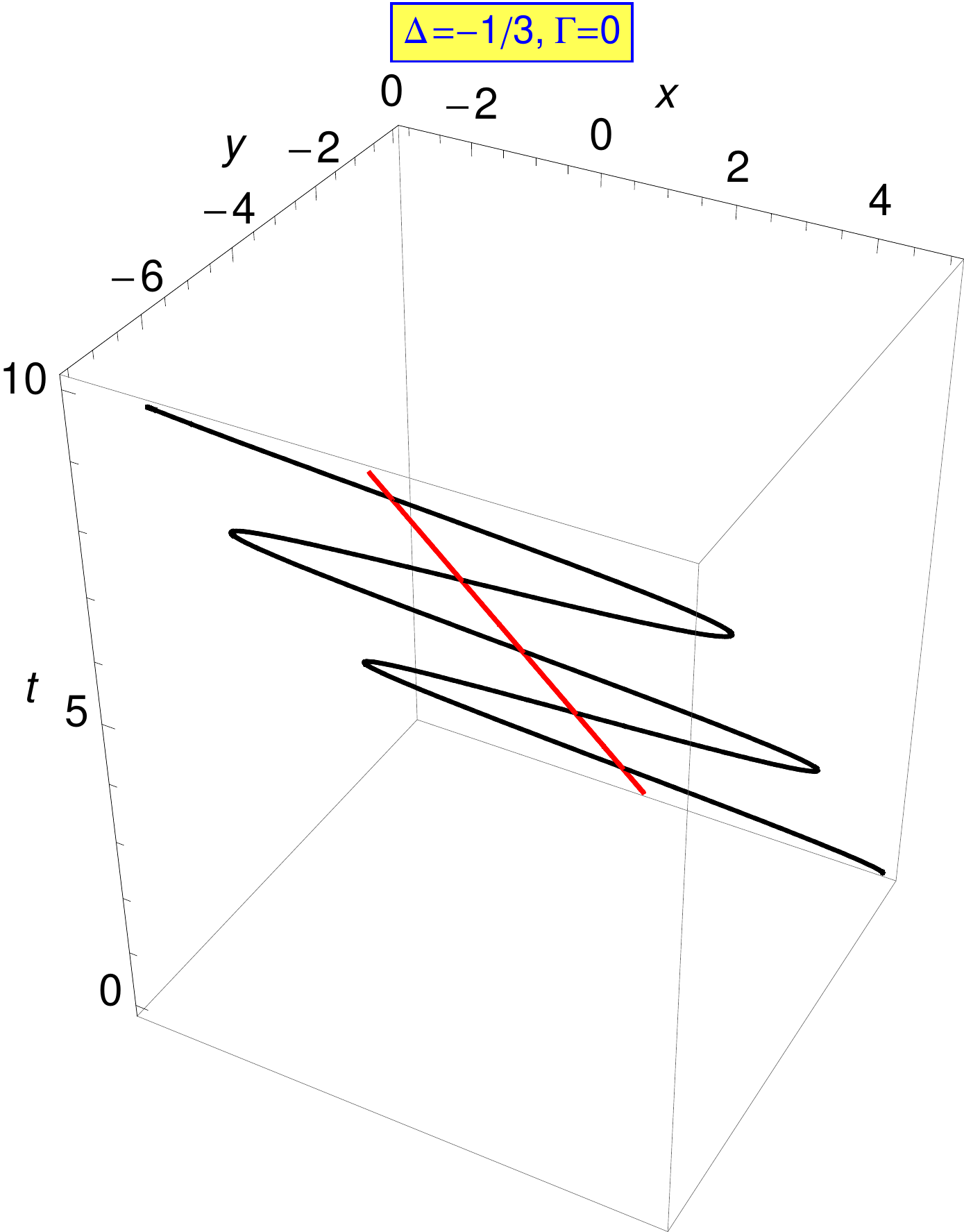}\\
\vspace{-5mm}
\end{center}
\caption{\it For $\Gamma=0$ the oscillations of
the $y$ coordinate disappear but $x(t)$ still oscillates around the Hall trajectory of the guiding center.}
\label{k4-figs}
\end{figure}

The particular case $\Gamma=0$ is rather harmless~: it simply
switches off the oscillations of the second coordinate of $\vX_+$, whereas its first
coordinate still oscillates, namely with frequency $\omega^*=-3\omega$.
Intuitively, we have sorts of ``half-Hall motions'',
see Fig. \ref{k4-figs}.
\goodbreak

In the critical case 
$
\Delta=0
$
i.e. for   angular velocity and radius
\beq
\omega=\omega_c=\frac{1}{2m\theta},
\quad\hbox{i.e.}\quad
R_c^3=4GMm^2\theta^2,
\label{CR}
\eeq
respectively,
the system is singular, and (\ref{X+-eq}) only allows
$ 
\vX_+(t)=0.
$ 
Thus the elliptic motions about the guiding 
center are eliminated, leaving us
with the guiding center motion alone, 
$ 
\vx(t)=\vX_-(t)
$ 
in (\ref{X-Hall}). Remarkably, the latter
 \emph{satisfies the Hall law}: the 
 center-of-mass moves in the planar ``magnetic'' field $eB=2m\omega$
perpendicularly to the ``electric'' field 
$e\vE=(3m\omega^2x_0,\,0)$ with appropriate ``Hall'' velocity
$E/B=3\omega x_0/2$.

As seen on Fig. \ref{k-figs}, when the critical value is approached, the rotation speeds
more and more up, and, \emph{with the exception of
those whose initial conditions are consistent with the Hall law},
all motions become ``instantaneous'' \cite{ZH-chiral}.

\section{Symmetries}

The same equations (\ref{HillSymmeqs}) also describe the
symmetries.
Having decomposed our  system into chiral components, the symmetry plainly follows from those of
our chiral solutions.
For the separated equations (\ref{X+-eq}) the initial conditions
$\vP=\vX_-(0)$ and $\vK=\vX_+(0)$,
\beqa
\vP&=&\barray{c}
x_0\\[6pt] y_0\earray
=
\barray{c}
X_-^1(t)
\\[6pt]
X_-^2(t)+\displaystyle\frac{3}{2}\omega tX_-^1(t)
\earray,
\\[10pt]
\vK&=&\barray{c}
-B/\omega^*
\\[6pt] 2\Gamma A/\omega^*
\earray=\nn
\\[6pt]
&&\barray{c}X_+^1(t)\cos\omega^* t-
\smallover{1}/{2\Gamma}X_+^2(t)\sin\omega^* t
\\[8pt]
 2\Gamma X_+^1(t)\sin\omega^* t
+X_+^2(t)\cos\omega^* t
\earray,
\eeqa
are plainly constants of the motion. They are interpreted as
translations and generalized boosts, respectively. 
For their Lie algebra structure we find, off the critical case $\Delta\neq0$, the commutation relations
\beqa
\{P^1, P^2\}=-\frac{2}{m\omega}\,,
\label{X-Halg}
\qquad
\{K^1, K^2\}=\frac{\Gamma}{\Delta}\frac{2}{m\omega}\,,
\label{X+Halg}
\eeqa
supplemented with $\{\vP, \vK\}=0$. We have therefore 
\emph{two uncoupled Heisenberg algebras
with {different} central charges $2/\omega$
and  $2\Delta/\omega$, respectively}.
Adding the obvious time-translation-symmetry would provide
us with  exotic Newton-Hooke symmetry without rotations
cf. \cite{ZGH}. 

In the critical case $\Delta=0$ the $\vX_+$ dynamics becomes trivial. $\vK=0$ drops out,
and the symmetry algebra reduces to the single Heisenberg algebra of $\vP=\vX_-$ alone [plus time translations], with extension parameter
determined by the non-commutative parameter,
$2/\omega_c=4m\theta$.  

\section{Conclusion}
%
Guided by  the analogy with the non-commutative Landau problem \cite{NCLandau,ZH-chiral},
we extended our previous study of Hill's equation to exotic particles. 
 Our most interesting result says that for a critical 
angular velocity i.e. for a critical radius
determined by the non-commutative parameter $\theta$,
cf. (\ref{CR}),
the only motions are those determined by the Hall law.
The role of $\theta$ is to enhance the ``Hall-type'' behavior, eliminating all the others in the critical case $\Delta=0$.
We note that Hall motions in stellar dynamics have been
considered before \cite{Binney,Chandra}.

Except for the lack of rotational symmetry due to the anisotropic oscillator term in (\ref{HillSymmeqs}),
our results are reminiscent to those for the
non-commutative Landau problem \cite{DHexo,NCLandau,AGKP,ZH-chiral}, and generalize those derived in \cite{ZGH}
for $\theta=0$. 

It is worth mentioning that the dimensional drop in
the critical case exemplifies the  degeneration
studied in \cite{Zanelli} in a general setting.

\begin{acknowledgments}
This paper grew out from joint work 
 with Gary Gibbons \cite{ZGH}, to whom 
we are indebted  for his interest and advice.
P.A.H  acknowledges  hospitality at the \textit{
Institute of Modern Physics} of the Lanzhou branch of
the Chinese Academy of Sciences. 
 This work was  partially supported by the National Natural Science Foundation of
China (Grant No. 11035006 and 11175215) and by the Chinese Academy of Sciences visiting
professorship for senior international scientists (Grant No. 2010TIJ06).
\end{acknowledgments}


\end{document}